\documentclass[12pt,thmsa]{article}
\usepackage{amssymb}

\usepackage{sw20lart}

\input{tcilatex}
\begin{document}

\title{Vacuum fluctuations the clue for a realistic interpretation of quantum
mechanics }
\author{Emilio Santos \\
Departamento de F\'{i}sica. Universidad de Cantabria. Santander. Spain}
\maketitle

\begin{abstract}
Arguments are gived for the plausibility that quantum mechanics is a
stochastic theory and that many quantum phenomena derive from the existence
of a real noise consisting of vacuum fluctuations of all fundamental fields
existing in nature. Planck\'{}s constant appears as the parameter fixing the
scale of the fluctuations. Hints for an intuitive explanation are offered
for some typical quantum features, like the uncertainty principle or the
particle behaviour of fields. It is proposed that the recent discovery of
dark energy in the universe is an argument for the reality of the vacuum
fluctuations. A study is made of the compatibility of the model with the
results of performed tests of Bell\'{}s inequalities.
\end{abstract}

\section{Introduction}

Understanding quantum mechanics presents big difficulties for many people, a
paradoxical fact in view of the relevance and the practical success of the
theory. For physicists interested in applications understanding is not very
relevant if it means getting an intuitive picture of the microworld. In
fact, for them the essential purpose of physics is to allow predicting the
results of experiments. Other physicists, including most of the workers in
foundations, think that the real problem is that, in the attempt to
understand quantum mechanics, we should not use concepts of classical physics%
\cite{Mittelstaedt}, or we should adhere to a kind of ``weak'' objectivity%
\cite{d'Espagnat}. In contrast with those opinions, in this article it is
proposed that quantum mechanics is less different from classical physics
than usually assumed, and it might be understood in a similar manner.

Quantum mechanics is extremely efficient for the prediction of experimental
results. In almost one century no significant violation of a quantum
prediction has been shown. Furthermore the agreement with experiments is
truly spectacular, reaching sometimes a precision of one part in $10^{10}.$
In contrast the interpretation of the quantum formalism has been the subject
of continuous debate since the very begining of the theory\cite{WZurek}
until today. This paradoxical situation, practical success combined with
conceptual difficulties, is something new in science. It is true that all
previous theories have given rise to controversy, but the conflict is more
accute in the case of quantum mechanics. An undesirable consequence has been
some confussion between science and pseudoscience in the public opinion. At
present some alleged consequences of quantum theory, like the uncertainty
principle or the impossible separation between object and subject, have
transcended the scientific community and are commented in newspapers and
popular writings, frequently presenting quantum mechanics like magic. The
situation has been originated, to some extent, by quantum physicists
themselves who have sometimes stressed the difficulties of understanding,
and even the wonder of that fact. In my opinion the lack of understanding is
not wonderful but unfortunate. In any case this state of affears does not
contribute to the popular esteem of true science.

``Any serious consideration of a physical theory must take into account the
distinction between the objective reality, which is independent of any
theory, and the physical concepts with which the theory operates. These
concepts are intended to correspond with the objective reality, and by means
of these concepts we picture this reality to ourselves''. With these words
begins the celebrated 1935 article by Einstein, Podolsky and Rosen\cite{EPR}$%
.$ The paragraph clearly supports a realistic interpretation of any physical
theory, which should be able to provide a model of the natural world.
Presenting a sketch of such a model for quantum theory is the purpose of
this paper. I shall not discuss the philosophical question of whether there
is an external world, independent of our mind, which is usually called
ontological realism. In any case I support \emph{epistemological realism,
that is the assumption that science makes assertions about the natural
world, }and not only about the results of observations or experiments\emph{.}

In my view there is not yet a clear physical model behind the quantum
formalism, is spite of the numerous interpretations proposed so far. In this
respect it is the opposite to general relativity. The latter provides a
clear, although strange, physical model: matter produces curvature of
spacetime and motion is governed by that curvature. However the
calculational tool (the Riemann geometry) is difficult to manage. In quantum
mechanics there is a beautiful and relatively simple formalism, that is
linear equations involving vectors and operators in a Hilbert space, but
there is no clear physical model behind. I would say that general relativity
has physical beauty, the quantum formalism has mathematical elegance.

Historically the renounciation to physical models in quantum mechanics came
as a consequence of frustration, due to the failure of the models proposed
during the first quarter of the 20th. century. This was the case after
Bohr's atomic model, consisting of point electrons moving in circular orbits
around the nucleus. That model, generalized with the inclusion of elliptical
orbits, produced some progress during the first few years after 1913.
However it was increasingly clear that the model was untenable. In 1926 an
alternative model was proposed by Schr\"{o}dinger, with an interpretation of
his wave mechanics where the electrons were considered continuous charge
distributions. As is well known that model was soon abandoned after the
correct criticisms by Bohr, Heisenberg and other people. Independently
Heisenberg had proposed a formalims, with the name of quantum mechanics,
which \emph{explicitly rejected any model}. Indeed he supported the view
that the absence of a picture was a progress towards a more refined form of
scientific knowledge. The success of the new quantum mechanics in the
quantitative interpretation of experiments, toghether with the failure to
find a good physical model of the microworld, led to the almost universal
acceptance of \emph{the current view that models are unnecessary or even
misleading}.

I do not agree with that wisdom, but support the EPR view\cite{EPR} that
there should be ``\emph{concepts intended to} \emph{correspond with the
objective reality, by means of which we picture the reality} \emph{to
ourselves}'' . That is any physical theory should contain two ingredients: a 
\emph{physical model} and a \emph{calculational tool, }the latter\emph{\ }%
including the formalism and rules for the connection with the experiments.%
\emph{\ }The calculational tool is essential because it is required for the
comparison of the theory with experiments. Indeed that comparison is the
test for the validity of the theory. However the physical model is necessary
in order to give satisfaction to the human being, who naturally aspires to
have a picture of the world. Furthermore the existence of a physical model
might open the possibility for new developments or applications of the
theory and therefore it is not a mere matter of taste.

In the next section I will sketch a physical model of the microworld which
is realist and clear and some specific quantum phenomena are analyzed within
that model. A warning is necessary. Very probably any quantum physicist
reading these two sections will quickly point out many contradictions
between the model and known empirical facts, so discarding the model as
untenable. I claim that the alleged contradictions are not real, they derive
from a common understanding of the quantum formalism which is not fully
correct. I will devote the third section of the article to rebutte the most
relevant of the apparent difficulties of the model, the alleged empirical
violations of Bell's inequalities. Finally I will point out briefly how
other difficulties might be solved.

\section{The physical model of the microworld}

\subsection{Vacuum fluctuations, the clue to understand quantum physics}

The concept of isolated system is the cornerstone of classical physics. But,
is isolation possible in our universe? I think not. The universe is not
static and so complex that some amount of noise is unavoidable. Furthermore,
the assumption of a fundamental noise may lead from classical physics to
quantum physics. Let us give an example, the stability of the hydrogen atom.
The atom consists of a proton (say at rest) and an electron moving around.
Everybody knows that a classical atom cannot be stable because the electron
would radiate, lossing energy and falling towards the proton. The argument
would be fine if there were an unique atom in space, but if there are many
atoms it is natural to assume that the radiation of one atom will eventually
arrive at other atoms. Thus every atom will sometimes emit radiation but
absorb it other times, possibly reaching a dynamical stationary state with
fluctuating energy. I shall not elaborate this example further, my purpose
being to convince the reader that the existence of a ``fundamental noise''
is plausible and might explain, at least, some phenomena taken as purely
quantal. Thus classical physics may be seen as an approximation of quantum
physics when the said noise may be neglected or, more properly, averaged out.

A fundamental noise is actually accepted in standard quantum theory, where
it is named ``quantum vacuum fluctuations''. These fluctuations are
qualified as ``virtual'', a word whose meaning is not clear. In my opinion
the acceptance of the quantum noise as real is what allows reaching an
intuitive picture of quantum physics

This leads me to the following picture of the quantum world. Fundamental
fermions, like leptons or quarks, are (localized) particles, but fundamental
bosons like photons, gluons, Z$_{0}$, W$^{\pm }$ or Higgs, are actually
(extended) fields. Gravity plays a special role, I support the view that
general relativity determines the structure of (curved) spacetime and its
relation with matter, so that gravity is not a field, at least not in the
same sense than the other fields. The wave behaviour of particles derives
from the unavoidable interaction with fields, and the particle behaviour of
fields derives from the interaction with particles, e. g. during detection.
A fundamental property of the universe is the existence of (vacuum, i. e.
not involving a finite temperature) fluctuations of all fields. In the case
of Bose fields there are random fluctuations similar to the zeropoint
fluctuations of the electromagnetic radiation to be studied below in some
detail. In the case of particles the fluctuations correspond to the
existence of a kind of ``Dirac sea'' of particles and antiparticles which
may be created and annihilated at random. There should be also metric
fluctuations of spacetime itself. Quantum commutation (anticommutation)
rules provide a disguised form of stating the properties of the fluctuations
in the case of fundamental fields (particles).

The existence of vacuum fluctuations gives rise to two characteristic traits
of quantum physics. Firstly quantum theory should be \emph{probabilistic}.
Secondly it should present a kind of \emph{wholeness}, quite strange to
classical physics where the concept of isolated system is crucial. The fact
that the vacuum fluctuations at different points may be \emph{correlated} is
the origin of the wholeness. Indeed \emph{entanglement }may be a consequence
of the correlation of quantum fluctuations at different points.\emph{\ }

\subsection{Planck's constant, the parameter fixing the scale of the
fluctuations}

For the sake of clarity let me consider just a kind of noise, the zeropoint
field fluctuations (ZPF) of the electromagnetic field. What are its
characteristics?. The most relevant is the \emph{spectrum}, which may be
defined as the energy density, $\rho \left( \nu \right) ,$ per unit
frequency interval, $d\nu $. It is remarkable that the spectrum is fully
fixed, except for a constant, by the condition of relativistic (Lorentz)
invariance. The direct proof is not difficult but I shall replace it by an
argument which may be traced back to Wien\'{}s work in 1894. An advantage of
that indirect derivation is that it discriminates clearly thermal noise from
the ZPF. Combining thermodynamics with Maxwell\'{}s electromagnetic theory
Wien derived the displacement law, which states that the spectrum of the
black body at a temperature $T$ should be of the form $\rho \left( \nu
,T\right) =\nu ^{3}f\left( \nu /T\right) .$ Lorentz invariance at zero
Kelvin is implicit in the use of electromagnetic theory. Now I claim that
putting $f\left( \nu /T\right) \rightarrow 0$ for $T\rightarrow 0$ leads to
classical physics, but putting the limit equal to a finite constant leads to
quantum physics. It is obvious that the constant involved should play a
fundamental role. It must be fixed by appeal to the experiments and the
result is that at $T\rightarrow 0$%
\begin{equation}
\rho \left( \nu \right) d\nu =\frac{4\pi }{c^{3}}h\nu ^{3}d\nu ,  \label{1.3}
\end{equation}
\emph{c} being the speed of light. Thus Planck\'{}s constant, $h,$ appears
with a transparent meaning, it fixes the scale of the \emph{universal noise}
or \emph{quantum noise (}but remember, I consider that the noise consists of%
\emph{\ real fluctuating fields}.) The ZPF spectrum eq.$\left( \ref{1.3}%
\right) $ corresponding to zero Kelvin, at a finite temperature the
fluctuating radiation contains a thermal part with Planck's spectrum. In
cosmology that part is called cosmic background radiation. Actually eq.$%
\left( \ref{1.3}\right) $ is strongly ultraviolet divergent, which posses a
big difficulty for the model here supported. In standard quantum field
theory the same difficulty appears, but it is eliminated by removing it via
the ``normal ordering rule'' in some cases and labelling it ``virtual'' when
a complete removal contradicts the observations. However this solution is
not good, it is better to assume that at high frequencies a cut-off should
exists due, for instance, to creation or annihilation of particles and/or to
gravitational (general relativistic) effects. I do not know how to implement
quantitatively this hypothesis but I will point out a possible solution to
the divergence problem later on (see section 2.7.)

The spectrum eq.$\left( \ref{1.3}\right) $ does not fully characterize the
ZPF. We must fix the joint probability distribution of the amplitudes of the
fields. A natural assumption is that the different modes are statistically
independent and each ones has a Gaussian distribution. Thus the ZPF becomes
a Lorentz invariant Gaussian stochastic field which may be characterized by
a plane waves expansion of the electric field 
\begin{equation}
\mathbf{E(r},t)=\frac{1}{\sqrt{V}}\sum_{j}\left[ c_{j}\mathbf{\varepsilon }%
_{j}\mathbf{(r})\exp \left( i\mathbf{k}_{j}\mathbf{.r}-i\omega _{j}t\right)
+c_{j}^{*}\mathbf{\varepsilon }_{j}^{*}\mathbf{(r})\exp \left( -i\mathbf{k}%
_{j}\mathbf{.r+}i\omega _{j}t\right) \right] ,  \label{2.1}
\end{equation}
where $\omega _{j}=2\pi \nu _{j}$, $\mathbf{\varepsilon }_{j}$ being
polarization vectors and $V$ a normalization volume. A similar expansion may
be written for the magnetic field with $i\mathbf{k.\varepsilon }_{j}$
substituted for $\mathbf{\varepsilon }_{j}$. Then the (complex) coefficients 
$\{c_{j}\}$\textbf{\ }and\textbf{\ }$\mathbf{\{}c_{j}^{*}\}$ form a set of
independent Gaussian random variables with zero mean and a square mean such
that eq.$\left( \ref{1.3}\right) $ holds true. That is, the probability
distribution of the coefficients will be 
\begin{eqnarray}
&&W\left( \{c_{j},c_{j^{*}}\}\right) d(\func{Re}c_{j})d(\func{Im}c_{j}) 
\nonumber \\
&=&\Pi _{j}\left[ \left( \pi 
\rlap{\protect\rule[1.1ex]{.325em}{.1ex}}h%
\omega _{j}/2\right) ^{-1}\exp \left( -\frac{2\left| c_{j}\right| ^{2}}{%
\rlap{\protect\rule[1.1ex]{.325em}{.1ex}}h%
\omega _{j}}\right) \right] d(\func{Re}c_{j})d(\func{Im}c_{j}).  \label{2.1a}
\end{eqnarray}

Up to here I have considered the electromagnetic field, but I will assume
that a similar ZPF exists for all fundamental Bose fields. Indeed all of
them should be in a dynamical equilibrium because they may interact
exchanging energy. The interaction will be stronger when the frequencies of
the excitations of the fields happen to have the same frequency. \emph{In
summary, a fundamental assumption of the physical model behind quantum
theory, }supported in this paper\emph{, is the existence of a (real)
universal noise, present even at zero Kelvin, consisting of Gaussian
fluctuations of all fundamental Bose fields of nature with an average energy 
}$\frac{1}{2}h\nu $\emph{\ for every normal mode (except at very high
frequencies.)}

A natural consequence of the belief in a real random electromagnetic
radiation with spectrum eq.$\left( \ref{1.3}\right) $ filling the whole
space has been the development of a theory known as \emph{stochastic (or
random) electrodynamics}. It deals with the study of electrically charged
particles immersed in the vacuum electromagnetic zeropoint field possessing
the spectrum eq.$\left( \ref{1.3}\right) $. The theory may be traced back to
Walter Nernst, who extended to the radiation field the 1912 second radiation
theory proposed by Planck for oscillators. That theory consisted of adding
the term $\frac{1}{2}h\nu $ to the oscillator energy at a finite temperature
thus giving the second Planck\'{}s law 
\[
U(\nu ,T)=\frac{1}{2}h\nu +\frac{h\nu }{\exp \left( h\nu /kT\right) -1}. 
\]
The idea was forgotten after the success of Bohr's atom, but it has been
rediscovered several times and developped by a small number of authors
during the last 50 years. A review of the work made until 1995 is the book
by de la Pe\~{n}a and Cetto\cite{dice}. Stochastic electrodynamics has
succeeded in providing a clear intuitive explanation for some phenomena
considered as purely quantal, and has been considered as a possible
alternative (or reinterpretation) of quantum mechanics by some authors. In
my opinion it is a semiclassical theory which reproduces quantum predictions
in a limited domain. In any case that theory has been the inspiration of the
physical model here presented.

It is here appropriate to skecth the explanation offered by stochastic
electrodynamics to the Casimir and Unruh effects\cite{dice} because both
reinforce the essential idea of this paper: that vacuum fluctuations are
real. The Unruh effect is the modification of the spectrum eq.$\left( \ref
{1.3}\right) $ when a radiation (classical) field is seen from an
accelerated reference frame. The Casimir effect corresponds to the fact that
the normal modes of the (classical) radiation are modified by the presence
of two metallic plates. If we ascribe an energy $\frac{1}{2}h\nu $ per
normal mode, the energy depends on the distance beween plates, leading to an
attraction between them which is the Casimir effect. It may be also seen as
due to the fact that the radiation pressure on the two sides of a plate is
no longer balanced when there is another parallel plate near it, due to the
modification of the normal modes.

\subsection{Heisenberg uncertainty relations}

A rather obvious consequence of the reality of the ZPF, proposed in this
paper, is that bodies cannot have smooth trajectories. In fact the forces
due to the vacuum ffluctuating fields give rise to a rapid random motion of
any particle. Thus the instantaneous velocity is meaningless, or irrelevant,
a situation similar to what happens in the theory of Brownian motion. Only
the mean velocity, $v$, during some time interval, $T$, may be a sensible
quantity. Still the mean velocity should fluctuate, with a fluctuation $%
\Delta v$ which would decrease with increasing time. Thus it is plausible to
assume that the product $T\Delta v$ should be a function of the variables
involved, that is the mass of the body, $m$, the Planck constant, $
\rlap{\protect\rule[1.1ex]{.325em}{.1ex}}h%
$, fixing the scale of the ZPF, and the velocity, $v,$ itself. Thus
dimensional considerations lead to 
\[
T\Delta v\simeq const.\frac{
\rlap{\protect\rule[1.1ex]{.325em}{.1ex}}h%
}{mv}\Rightarrow Tv\Delta v=\Delta x\Delta \nu \simeq \frac{%
\rlap{\protect\rule[1.1ex]{.325em}{.1ex}}h%
}{2m}, 
\]
where $\Delta x$ is the distance traveled by the body during the time $T$,
and I have chosen the constant in order to agree with the quantum
prediction. In summary, due to the vacuum fluctuations, smooth paths do not
exist and it is not possible to define both the position and the velocity of
a particle with precision beyond the one allowed by the Heisenberg
uncertainty relation.

I must point out that the comparison with Brownian motion may be misleading
due to the big difference between their spectra. This may be better seen
from the selfcorrelation function, which is the Fourier transform of the
spectrum, that is 
\begin{equation}
\left\langle x(0)x(t)\right\rangle =\int_{0}^{\infty }S\left( \omega \right)
\cos \left( \omega t\right) d\omega .  \label{corr}
\end{equation}
In Brownian motion the spectrum is white (i.e. a constant) so that the
integral eq.$\left( \ref{corr}\right) $ is zero. This means that the
Brownian particle losses quickly any memory of the initial position. In
contrast the spectrum of the ZPF increases rapidly, it being zero for $%
\omega \rightarrow 0$ (see eq.$\left( \ref{1.3}\right) $). As a consequence
the cosinus term effectively cuts-off the spectrum for $\omega \gtrsim 1/t,$
but some correlation remains. This means that the memory of the initial
position is not fully lost. The picture that emerges is that particles have
highly irregular paths which may be seen as the superposition of a smooth
one plus strong short-time fluctuations. Actually the effect of the ZPF
fluctuations decreases with the particle\'{}s mass so that the classical
limit (absence of fluctuations) is obtained for very massive bodies.

Another form of the uncertainty relation appears in a stationary motion. Let
us consider a particle in an external potential well. Due to the existence
of the ZPF, the particle will perform a more or less periodic motion with a
random frequency, say of order $\nu .$ (If the particle is charged the
electromagnetic ZPF will be relevant field, but in any case the metric
fluctuations will play a fundamental role.) As the ZPF has energy $\frac{1}{2%
}h\nu $\emph{\ }per normal mode of the radiation, we may assume that the
particle reaches a dynamical equilibrium with the ZPF when its kinetic
energy is similar to half the energy of the appropriate mode, that is 
\begin{equation}
\frac{1}{2}mv^{2}\approx \frac{1}{4}h\nu .  \label{2.4}
\end{equation}
It is possible to get a relation independent of the frequency because the
mean square velocity may be related to the mean square displacement via 
\begin{equation}
\left\langle v_{x}^{2}\right\rangle \approx 4\pi ^{2}\nu ^{2}\left\langle
\left( x-\left\langle x\right\rangle \right) \right\rangle ^{2}\equiv 4\pi
^{2}\nu ^{2}\Delta x^{2}.  \label{2.4a}
\end{equation}
Hence, taking eq.$\left( \ref{2.4}\right) $ into account we get 
\begin{equation}
\Delta x\Delta p_{x}\simeq \frac{
\rlap{\protect\rule[1.1ex]{.325em}{.1ex}}h%
}{2},
\rlap{\protect\rule[1.1ex]{.325em}{.1ex}}h%
\equiv \frac{h}{2\pi },\Delta p_{x}^{2}\equiv m^{2}\left\langle
v_{x}^{2}\right\rangle .  \label{2.5}
\end{equation}

This provides an intuitive interpretation of the Heisenberg uncertainty
relations as follows.\textit{\ Due to the quantum noise (vacuum
fluctuations) with the peculiar spectrum eq.}$\left( \ref{2.1}\right) ,$ 
\emph{1)} \emph{the paths of particles are not smooth but fluctuating, so
that velocity and position cannot be defined simultaneously with arbitrary
precision, and 2) }\textit{it is imposible to localize a particle in a
region of size }$\Delta x$\textit{\ without the particle having a random
motion with typical momentum dispersion }$\Delta p\gtrsim 
\rlap{\protect\rule[1.1ex]{.325em}{.1ex}}h%
/2\Delta x.$ Thus the uncertainty relation appears as a practical limit to
the localization of particles in phase space, rather than a fundamental
principle of ``uncertainty''. However in practice the difference is less
relevant than it may appear. For instance as all measuring devices are
immersed in the ZPF, the interaction of with a microscopic system gives a
random character which necessarily leads to a ``disturbance induced by the
measurement''. This fact may explain the ``Heisenberg microscope'' and other
effects associated to the uncertainty relations.

\subsection{A picture of molecules and solids}

As is well known the Heisenberg relations allow estimating the size and
energy of the ground state of any quantum system. Thus these properties may
be interpreted intuitively as due to the fact that all systems are immersed
in the universal quantum noise (the vacuum fluctuations.) This suggests a
picture of quantum systems which is lacking in standard quantum mechanics.
For example, in the ground state of the hydrogen atom quantum mechanics
attributes to the electron a well known spherically symmetric probability
distribution of positions, but the standard interpretation does not asserts
that the electron has an actual position at any time. It only states that
``if we performed a measurement of position we would obtain a specific value
with some probabilty predicted by quantum mechanics''. A picture of the
reality is therefore lacking. In contrast we offer a clear picture of the
electron performing a random motion. The picture may be extended to other
atoms provided we add two important ingredients: the spin and the Pauli
principle, but I shall not comment on them in this paper.

Let us now explain the model\'{}s picture of molecules. For the sake of
clarity I will consider as an example the $CO_{2}$ molecule. As is well
known it consists of three atoms in a stright line with the carbon atom at
the center and some precisely measured carbon-oxigen distances. Now quantum
theory states that the ground state of the isolated molecule possesses zero
angular momentum (say, making a calculation within non-relativistic quantum
mechanics, involving the electrons and three spin-zero nuclei taken as point
particles.) Indeed the molecule consists of three spin-zero nuclei plus an
even number of spin-1/2 electrons coupled to zero total spin angular
momentum. Thus a quantum calculation leads to the prediction that the ground
state has zero total angular momentum. Now zero angular momentum implies
invariance to rotations, that is spherical symmetry. How may we understand
the contradictory facts that the molecule is linear and it has spherical
symmetry?. The standard wisdom is that the linear form of the molecule \emph{%
emerges }only as a result of a measurement or in general by the interaction
with the environment when the molecule is not isolated. This interpretation
is rather counterintuitive, to say the least. In contrast our model offers a
transparent picture: the molecule retains its linear form at any time (aside
from vibrations,) but changes the orientation randomly with a spherically
symmetric probability distribution. This is similar to our model\'{}s
picture of the random motion of the electron in the hydrogen atom commented
above. In my opinion this is a case where the standard interpretation of the
quantum formalism should be reinterpreted. Indeed I propose that the quantum
prediction of a dispersion-free zero angular momentum actually means that
the dispersion cannot be detected by measurements, but we should not believe
that no rotation exists. This is an example of the apparently unsurmontable
difficulties for a physical model of the quantum world.

\subsection{Particle behaviour of fields}

I assume that in nature there are particles and fields (waves). In our model
the fundamental Bosons (Fermions) are fields (particles), but composite
systems like baryons, nuclei, atoms or molecules, are particles either if
they possess integer or half-odd angular momentum. There is no problem to
understand the localized detection of particles or the interference of
waves, but there are difficulties to get a picture of the wave behaviour of
particles or the corpuscular behaviour of waves. Here some hints will be
provided for an intuitive understanding of that behaviour.

I shall start with the corpuscular behaviour of fields, discussing only the
electromagnetic field although the arguments may be general.The first
problem is why the ZPF, which is very strong at high frequencies, does not
activate light detectors. The obvious solution is to postulate that \emph{%
detectors are activated only by the radiation in excess of the ZPF}. This
assumption is actually the same postulated in standard quantum theory.
Indeed the ``normal ordering'' rule makes exactly a subtraction of the ZPF.
For instance if we made a quantum calculation of the vacuum expectation of
the energy density, $\rho _{vac}$, without the normal ordering rule we would
get 
\begin{equation}
\rho _{vac}^{(wrong)}=\sum_{j}\langle vac\left| H_{j}^{(wrong)}\right|
vac\rangle =\frac{h}{2}\sum_{j}\nu _{j}\langle vac\left| (a_{j}+a_{j}\dagger
)^{2}\right| vac\rangle =\sum_{j}\left( \frac{1}{2}h\nu _{j}\right) ,
\label{rovac}
\end{equation}
where $a_{j}$ ($a_{j}\dagger )$ is the annihilation (creation) operator of
photons. In the limit of large normalization volume the sum in $j$ becomes a
frequency integral of eq.$\left( \ref{1.3}\right) ,$ which is clearly
divergent. The quantum solution to this problem is to modify the Hamiltonian
(and more generally all expresions quadratic in the fields) postulating the
``normal ordering rule'', which means putting the annihilation operators to
the right. Thus the quantum calculation gives 
\begin{eqnarray*}
\rho _{vac} &=&\sum_{j}\langle vac\left| H_{j}\right| vac\rangle =\frac{h}{2}%
\sum_{j}\nu _{j}\langle vac\left| 2a_{j}\dagger a_{j}\right| vac\rangle \\
&=&\frac{h}{2}\sum_{j}\nu _{j}\langle vac\left| (a_{j}+a_{j}\dagger
)^{2}\right| vac\rangle -\sum_{j}\left( \frac{1}{2}h\nu _{j}\right) =0,
\end{eqnarray*}
where I have taken into account the standard commutation rules. It may be
realized that passing to the normal ordering is equivalent to subtracting
the ZPF contribution eq.$\left( \ref{rovac}\right) .$

In formal quantum mechanics the normal ordering rule may be accepted as a
postulate without any further discussion. In contrast if we want to find a
physical model behind the formalism, the rule of subtracting the ZPF
presents difficulties in the case of photon counters. Remember that in our
model there are no ``particles of light'' (photons), but light consists of a
continuous radiation, always including the ZPF. The problem is that in
practice the ZPF cannot be subtracted exactly. This fact does not derive
from the (divergent) energy associated to the ZPF, but from the fact that
the ZPF is fluctuating. If we assume that the ZPF is real, there is always
the possibility that some vacuum fluctuations are confussed with signals by
the detectors, no matter how efficient is the subtraction made. This is
explained in the following, where I study three relevant kinds of light
detectors.

Firstly we may consider bolometric detection, which is frequently used in
astronomy. It consists of a lens system which collects all light arriving at
it during some time and it is measured the total energy stored. In this case
the ZPF subtraction may be quite accurate because the fluctuations average
out if the measurement time is long enough.

The detection of ``individual photons'' in a photographic plate is due to
the atomic nature of the plate. In this case saying that radiation are
particles because they give rise to individual blackened grains is like
saying that wind is corpuscular because the number of trees falling in the
forest is an integer. Of course in both cases, the photo and the forest,
there is a random element. It is obvious for the wind but, as explained
above, there is also a random element in the radiation: the quantum noise or
ZPF.

The detection process in a photon counter may be explained as follows.
Inside the detector there are systems, e. g. molecules, in a metastable
state. The arriving radiation, with a random element due to the ZPF, has
from time to time sufficient intensity to stimulate the decay of the
metastable system and this gives rise to a photocount. However the noise
alone, being fluctuating, may eventually produce counts in the absence of
any signal, which are called dark counts. (Dark counts are usually
attributed to thermal fluctuations, but I claim that quantum fluctuations
may produce a fraction of them.) The counter behaves like an alarm system.
If it has low sensitivity it may fail to detect some relevant signals, but
if it is too sensitive it may be activated by accident. The same is likely
true for photon counters. This leads me to conjecture that it is not
possible to manufacture detectors with 100\% efficiency but no dark counts
and that this trade-off is the origin of the socalled \emph{detection
loophole in the optical tests of Bell}$\acute{}$\emph{s inequalities }(see
section 3.4 below.) In any case explaining in detail how a detector may
subtract efficiently the ZPF is not trivial (it has been called ``the needle
in the haystack'' problem) and naive models predict too many dark counts\cite
{santossubtraction}.

There are many more phenomena where light appears as consisting of particles
(photons), but might be qualitatively interpreted in terms of waves. I shall
mention just two: the atomic emission in the form of needels of radiation
and the anticorrelation after a beam splitter.

Firstly I point out that the spontaneous emission of photons by atoms is
interpreted, in our model, as an emission estimulated by the ZPF. As a
simplified model we may consider that a plane wave with wavevector $\mathbf{k%
}$\textbf{, }belonging to the ZPF\textbf{, }arrives at an atom in the
appropriate conditions to emit radiation with angular frequency $\omega
=c\left| \mathbf{k}\right| $. If the radiation emitted is in the form of a
spherical wave, far from the atom we will observe the interference between
the incoming ZPF plane wave and the emitted spherical wave. It is not
difficult to realize that the interference will be constructive mainly
within an angle $\theta $ with respect to the forward direction $\mathbf{k,}$
such that 
\[
\frac{l}{\cos \theta }-l\sim \frac{\lambda }{2}, 
\]
where $l$ is the distance from the atom to the observation point and $%
\lambda =1/$ $\left| \mathbf{k}\right| .$ Thus the emitted radiation will
look like a needle having a width about $l\sin \theta \sim \sqrt{l\lambda }%
<<l,$ superimposed to the ZPF. That needle is what we may interpret as one
photon. Of course that simplified model does not provide an explanation as
to why the total energy of the photon is related to the frequency via $E= 
\rlap{\protect\rule[1.1ex]{.325em}{.1ex}}h%
\omega .$ This would require a detailed study of the structure of the atom
and its interaction with light. In any case the needle of radiation should
have a well defined energy, $
\rlap{\protect\rule[1.1ex]{.325em}{.1ex}}h%
\omega ,$ and momentum, $
\rlap{\protect\rule[1.1ex]{.325em}{.1ex}}h%
\omega /c.$ After that the Compton effect might be understood as a collision
between an electron and a needle of radiation$.$

The anticorrelation after a beam splitter was first empirically proved by
Grangier et al.\cite{Grangier}. The experiment essentially consists of
sending individual photons to a balanced beam splitter with two photon
counters placed after the two output ports of the splitter. The authors
observed that either a count is produced in the first detector or in the
second one, but never in both (most times no count was shown in either
detector because the efficiency was rather low). This is interpreted as
evidence for the particle behaviour of photons, which apparently cross the
splitter undivided.

An analysis of that result within our model is somewhat involved\cite{MS},%
\cite{SO} but a simplified explanation is as follows. In the input port of
the splitter not only the wavepacket (representing) the photon enters, there
is also an incoming radiation from the ZPF. We may represent the total
electric field at the entrance by the complex amplitude $E_{signal}+E_{ZPF}$%
, where I take into account that we should add the amplitudes of signal and
ZPF, not the intensities. There is also another input port of the splitter
where no signal enters, but some radiation from the ZPF should also enter,
which we may represent by $E_{ZPF}^{\prime }.$ The fields in the output
ports will correspond to appropriate additions of the incoming amplitudes,
that is 
\begin{equation}
\frac{1}{\sqrt{2}}\left( E_{signal}+E_{ZPF}\right) +\frac{i}{\sqrt{2}}%
E_{ZPF}^{\prime },\frac{i}{\sqrt{2}}\left( E_{signal}+E_{ZPF}\right) +\frac{1%
}{\sqrt{2}}E_{ZPF}^{\prime },  \label{2.2a}
\end{equation}
where $i\equiv \sqrt{-1}$ and I have taken into account that the transmitted
(reflected) radiation does not change the phase (change by $\pi /2.)$ Thus
the corresponding intensities arriving at the detectors will be 
\begin{equation}
I_{\pm }=\frac{1}{2}\left| E_{signal}+E_{ZPF}\right| ^{2}+\frac{1}{2}\left|
E_{ZPF}^{\prime }\right| ^{2}\pm \func{Im}\left[ \left(
E_{signal}+E_{ZPF}\right) ^{*}E_{ZPF}^{\prime }\right] .  \label{2.2}
\end{equation}
Actually only the part of the ZPF in the same mode of the signal (with the
same frequency) may interfere with it, and therefore that is the relevant
part. Also the corresponding ZPF intensity $(\frac{1}{2}h\nu )$ is half the
intensity of the signal $(h\nu ).$ Thus eq.$\left( \ref{2.2}\right) $ might
be rewritten, after subtraction of the ZPF intensity $I_{0},$%
\[
I_{\pm }-I_{o}=I_{signal}\left[ \frac{3}{4}+\frac{\sqrt{2}}{2}\cos \phi
_{1}\pm \left( \sqrt{2}\sin \phi _{1}+\sin \phi _{2}\right) \right] , 
\]
where $\phi _{1}\left( \phi _{2}\right) $ is the relative phase between $%
E_{signal}$ and $E_{ZPF}$ $\left( E_{ZPF}^{\prime }\right) ,$ all angles
assumed equally probable. The relevant point is that, except for some values
of the $\phi _{1}$ and $\phi _{1}$, not very probable, the intensity
arriving at one of the detectors is above the ZPF level (therefore
detectable), and below the ZPF level (therefore not detectable) at the other
detector, which may explain the anticorrelated detection. The full absence
of anticorrelation might be explained assuming that for low efficiency
detectors (like those used in the commented experiment\cite{Grangier}) only
radiation \emph{well above} the ZPF intensity would be detected.

\subsection{Wave behaviour of particles}

There is wide empirical evidence for a wave behaviour of particles. Asides
from the early evidence for electron difraction, the experimental results of
neutron interference are really impresive\cite{Rauch} and to a lesser degree
the atom interference. On the other hand the alleged interference of big
molecules, like fullerene, may be less significant\cite{Sulc}.

It is attractive the hypothesis that the wave behaviour of particles derives
from the interplay of the particles and the quantum noise. For instance, in
the interference of charged particles we might assume that the
electromagnetic ZPF in a periodic arrangement interferes with itself, giving
rise to maxima and minima of field intensity in some points, which might
guide the particle to the screen where the interference pattern is
exhibited. The picture has some similarity with the hypothesis, put forward
by de Broglie, that any particle is always accompanied by an associated
wave. L. de Broglie's proposal is usually understood as if every particle
possesses its own wave, a picture reinforced by the quantitative relation
between the particle's momenum, \textbf{p}, and the wavevector, \textbf{k}.
However that picture is untenable. For instance, how the (extended) wave may
follow the (localized) particle during the motion of the latter? It is more
plausible to assume that there is some background of waves in space able to
interact with the particles. This leads to the picture that the waves are,
actually, those of the quantum noise or ZPF.

The problem is to explain why the overhelming interaction of the particle
occurs with just one mode of the radiation, that is the one given by de
Broglie's relation. A number of people, in particular this author, have
attempted to develop quantitatively that model without real progress till
now.

\subsection{Dark energy}

The observed accelerated expansion of the universe is currently assumed to
derive from a positive mass density and a negative pressure, constant
throughout space and time, which are popularly known as ``dark energy''. The
mass density, $\rho _{DE},$ and the presure, $p_{DE},$ fitting the
observations are 
\begin{equation}
\rho _{DE}\simeq -p_{DE}\simeq 10^{-26}\text{ kg/m}^{3}.  \label{1}
\end{equation}
Many proposals have been made for the origin of dark energy, the most
popular being to identify it with the cosmological constant introduced by
Einstein in 1917 or, what is equivalent in practice, to assume that it
derives from the quantum vacuum. Indeed the equality $\rho _{DE}=-p_{DE}$ is
appropriate for the vacuum (in Minkowski space, or when the spacetime
curvature is small) because it is Lorentz invariant.

A problem appears because, if the dark energy is due to the quantum vacuum,
it should be either strictly zero or of order Planck\'{}s density, that is
about 123 orders of magnitude larger than eq.$\left( \ref{1}\right) $. In
particular the Planck density is roughly obtained if we integrate the energy
density of the electromagnetic ZPF, eq.$\left( \ref{1.3}\text{ }\right) ,$
up to Planck\'{}s frequency (the inverse of Planck\'{}s time.) Thus the
hypothesis that the ZPF vacuum energy is real seems flawed, which would make
the model of this paper untenable. However there is an alternative which
fits in the normal ordering rule of quantum field theory, equivalent to the
ZPF energy subtraction as discussed in section 2.5. We might assume that the
vacuum fluctuations of fundamental Bose fields, like the electromagnetic
radiation, contribute a positive energy, but fluctuations of fundamental
Fermi fields contribute a negative energy, in such a way that the total
vacuum energy is strictly zero. A strict cancelation would be quite
satisfactory and would justify the subtractions involved in the normal
ordering rule.

Nevertheless, even if the mean energy density is zero, as the fields are
fluctuating the square mean energy density \emph{must} be positive (not
zero). Indeed for a random variable $x$ the equalties $\left\langle
x\right\rangle =0=\left\langle x^{2}\right\rangle $ would imply the total
absence of fluctuations. The vacuum fluctuations should create a
gravitational field. Long ago Zeldovich\cite{Zel} proposed that this
mechanism may be the origin of the dark energy and estimated the
gravitational energy density to be 
\begin{equation}
\left| \rho _{DE}\right| c^{2}\sim G\frac{m^{6}c^{2}}{h^{4}}=\frac{Gm^{2}}{%
\lambda }\times \frac{1}{\lambda ^{3}},\lambda \equiv \frac{h}{mc}.
\label{40}
\end{equation}
This corresponds to assuming that fluctuations with typical mass $m$ take
place with a (positive) correlation length $\lambda .$ In fact the observed
value $\rho _{DE}$, eq.$\left( \ref{1}\right) ,$ is obtained if the mass $m$
is close to the pion mass. The sign of $\rho _{DE}$ in eq.$\left( \ref{40}%
\right) $ might be negative or positive depending on details of the
two-point correlation of the vacuum fluctuations.

The Zeldovich calculation of the gravitational energy, eq.$\left( \ref{40}%
\right) $ follows from Newtonian gravity. An explicit calculation within
general relativity\cite{darkenergy} shows that vacuum fluctuations lead to a
curvature of spacetime equivalent to the one which would be produced by a 
\emph{positive} energy and a \emph{negative} pressure as in eq.$\left( \ref
{1}\right) $. That calculation involves a single free parameter, the mass 
\emph{m} as in eq.$\left( \ref{40}\right) .$ An argument for the value of $m$
follows. We might especulate that the important fluctuations would be those
involving hadrons, produced by strong forces, and they would be more
probable if the particles created are light, so requiring less energy. This
would lead us to assume that the most relevant fluctuations might be those
involving the lightest hadron, that is the pion. In fact the pion mass $m$
in eq.$\left( \ref{40}\right) $ fits fairly well the observed dark energy
value eq.$\left( \ref{1}\right) .$

\section{Quantum mechanics vs. local realism. The Bell theorem}

\subsection{Are the laws of nature causal?. The debate about completeness}

There are strong difficulties to reach a physical model of the microworld
from the quantum formalism. Possibly the most relevant are the wave-particle
behaviour, above commented, and the violation of the Bell inequalities to be
studied below. These difficulties apparently prevent an intuitive picture of
the quantum world, e.g. the one provided by our model. But it is a fact that
these difficulties have been greatly enhanced by sociological and historical
reasons, as I comment in the following.

The early interpretation of quantum mechanics grew on the ground of the
philosophical doctrines of positivism and pragmatism. In the positivistic
view, physics should lie close to the empirical results, avoiding models
which might be wrong. For instance Heisenberg early quantum mechanics
rejected models of atoms from the start, using purely mathematical objects
(matrices) instead. On the other hand pragmatism states that, the prediction
of the experimental results being the unique criterion for the validity of a
theory, physical models are not necessary at all. Nevertheless I do not
claim that the positivistic and/or pragmatic attitutes were philosophical
prejudices without support in the empirical facts. I am aware that people
adhered to these doctrines due to the difficulties to understand atomic
physics with the realistic epistemology inherited from 19th century physics.
In particular, the lack of progress of the old quantum theory resting upon
Bohr\'{}s atomic\emph{\ model}.

The said philosophical background led Bohr to support the completeness of
quantum mechanics, that is to reject the necessity of a more detailed theory
able to offer a definite world view. Most people adhered to that view. A
consequence of the completeness assumption is that two states of a physical
system represented by the same statevector should be identical. But this
leads to the unavoidable conclusion that quantum probabilities derive from a
lack of strict causality of the natural laws. That is we should assume that
different effects may follow from the same cause. For instance two
radioactive atoms in the same excited quantum state decay at different
times, in spite of been identical according to the formalism. This is
usually called the \emph{fundamental or essential probabilistic} character
of the physical laws. Einstein disliked that assumption and strongly
criticized it, as summarized in his celebrated sentence ``\emph{God does not
play dice}''. The alternative supported by Einstein\cite{EPR}, \cite
{Einstein} (and other founding fathers like Schr\"{o}dinger\cite{Schrödinger}%
) is that quantum mechanics is not complete. That is the laws of nature are
strictly causal, but there is a random element, not explicit in the quantum
formalism, giving rise to the statistical character of quantum predictions.
According to this realistic approach two atoms in the same quantum state are
not identical, what is identical is our (incomplete) information about their
real states. This situation led to the well known debate about completeness,
confronting Bohr and Einstein positions. The model presented in this paper
obviously conforms to Einstein's view. The debate remained at the
philosophical level because it was assumed that strict causality combined
with randomness is in practice indistinguishable from essential probability.
However a possible empirical discrimination appeared after the work of John
Bell, to be commented below.

In my view it is not plausible to assume that a system can be fully isolated
from the rest of the world, because the vacuum fields may provide an
effective interaction with many other systems. However in order to be able
to make physics we should assume that microscopic systems, even if not
isolated, may be treated with a formalism which in some form takes account
of the interaction with the vacuum fields. For instance, if we represent the
state of an atom by a statevector, it is plausible to assume that this
representation corresponds to the atom ``dressed'' by all fields which
interact with it. As is well known in quantum electrodynamics the physical
electrons are never ``bare'' but ``dressed with virtual photons and
electron-positron pairs''. The word ``virtual'' is just a name for something
which has observable effects but is unknown. One of the aims of the model
here presented is to give a more clear meaning to the word virtual. One of
the reasons for the difficulties to understand quantum mechanics is that it
deals with a world where noise (randomness) is crucial, but the noise is
hidden because the Schr\"{o}dinger equation is deterministic (probabilities
appear only as a result of the measurement.)

Actually to pretend that a statevector represents faithfully the \emph{%
actual state of an individual system} is a rather presumptuous attitude. It
is more plausible to assume that the statevector represents the relevant
information available about the system, which gives support to the belief in
the incompleteness of quantum mechanics. This is the idea behind the
ensemble interpretation of quantum mechanics: the state vector should be
associated to a statistical ensemble of systems rather than to a single
systgem\cite{EPR}, \cite{Ballentine}\emph{. } In summary \emph{the
complexity of even the most elementary quantum system, like a ``dressed
electron'', makes the ensemble interpretation of the statevector most
plausible.}

\subsection{Hidden variables}

If we accept the ensemble interpretation the obvious question is the
following: In view that the information provided by quantum mechanics is
incomplete, should we attempt to get additional information?. If the answer
is \emph{yes} we should search for a subquantum level which could increase
our understanding of the world. That line of research is usually known as
the \emph{hidden variables programme}. But the answer may be \emph{not, }%
because the experience shows that the programme has failed in spite of the
big effort of some (few certainly) people during almost one century. In any
case it is my opinion that the rejection of the hidden variables programme
is not a sober attitude.

As said above the mainstream of the scientific community has been positioned
for the completeness of quantum mechanics and therefore against hidden
variables (HV). Possibly the origin of this fact lies in the strong
personality of Bohr combined with the confort produced by the belief that
one possesses the final theoretical framework of physics, that is quantum
mechanics. With the time the rejection to HV theories was reinforced by the
failure to find a useful one. In any case a strong influence had the
celebrated von Neumann's 1932 theorem against hidden variables\cite{von
Neumann}, which prevented the research on the subject during more than three
decades. In 1965 Bell\cite{BellRMP} showed that the physical assumptions of
von Neumann were too restritive and that (contextual) hidden variables are
possible. Indeed for any experiment it is a simple matter to find a specific
contextual hidden variables theory\cite{Scontext}. What is difficult is to
get a HV theory valid for the whole of quantum phenomena. The relevant point
is that the mere existence of HV has consequences that might be tested
empirically. In the following subsection I will derive the most relevant
ones following Bell\'{}s work\cite{Bell}.

The existence of hidden variables is related to philosophical realism, that
is the assumption that systems possess \emph{elements or reality}\cite{EPR}
(ontological realism) and that physics should make assertions about that
reality and not only about the results of the experiments (epistemological
realism.) A necessary condition for realism may be stated as follows. Let us
assume that in some experiment we want to measure the observable $A$. Then
the value, say $a$, obtained in a measurement will depend on the state, say $%
\lambda ,$ of the system and the observable which we measure. We should
write 
\begin{equation}
a=a\left( \lambda ,context(A)\right) .  \label{5}
\end{equation}
where we may interpret $\lambda $ as the elements or reality, i. e. the set
of values of the (maybe hidden) variables which faithfully determine the
state of the system. The dependence on \emph{context(A)} takes into account
that the result of the measurement may depend on the full experimental
equipment used for the measurement of the observable $A$. For some people eq.%
$\left( \ref{5}\right) $ is a condition for determinism\cite{Gisin}, rather
than realism, because the states of the system and the context completely 
\emph{determine }the result of the measurement. It is true that eq.$\left( 
\ref{5}\right) $ excludes the possibility that natural laws are not strictly
causal, and therefore the name causality would be more appropriate than
determinism, but I shall not discuss this semantical point further. In any
case we may replace eq.$\left( \ref{5}\right) $ by the more general one

\begin{equation}
\Pr (a)=P_{a}\left( \lambda ,context(A)\right) ,  \label{5a}
\end{equation}
with the meaning that the states of system and context only determine the
probability of getting the value $a$. Thus eq.$\left( \ref{5a}\right) $ is
compatible with both the assumption that natural laws are strictly causal
and its denial. We see that realism (whose necessary condition is eq.$\left( 
\ref{5a}\right) )$ is more general than determinism (or deterministic
realism, or strict causality) whose necessary condition is eq.$\left( \ref{5}%
\right) .$

The above construction allows defining two possible kinds of hidden
variables theories. In fact let us assume that we measure not one but two
observables, $A$ and $B$, in the same experiment. Thus eq.$\left( \ref{5a}%
\right) $ leads to the following expectation value for the product of the
two observables 
\begin{equation}
\left\langle AB\right\rangle =\int \rho \left( \lambda \right) d\lambda
\sum_{a}aP_{a}\left( \lambda ,context(A,B)\right) \sum_{b}bP_{b}\left(
\lambda ,context(A,B)\right) ,  \label{6}
\end{equation}
where I have assumed one continuous hidden variable with probability density 
$\rho \left( \lambda \right) $. As above the dependence on $context(A,B)$
takes into account that the results of the measurement might depend on the
full experimental context of the measurement. In other words, we do not
exclude that the expectation of the product of two observables, $%
\left\langle AB\right\rangle ,$ might lead to a different numerical value if
measured with a different equipment. With an assumption as general as eq.$%
\left( \ref{6}\right) $ it is not strange that, choosing appropriately the
functions $P_{a}\left( \lambda ,context(A,B)\right) $ and $P_{b}\left(
\lambda ,context(A,B)\right) ,$ we may reproduce any desired result for $%
\left\langle AB\right\rangle ,$ in particular a result in agreement with the
quantum prediction. HV theories where expectations are given by eq.$\left( 
\ref{6}\right) $ (or its generalization for more than two observables) are
usually named \emph{contextual}, but a more appropriate name would be \emph{%
general }HV theories. They are obviously compatible with quantum mechanics,
but being so general they are not too interesting.

A restricted family of HV theories consists of those \emph{non-contextual, }%
where the value of the observable $A$ does not depend on the whole
experimental context but only on the state of the system under study, and
similar for $B.$ In non-contextual HV theories eq.$\left( \ref{6}\right) $
is replaced by 
\begin{equation}
\left\langle AB\right\rangle =\int \rho \left( \lambda \right) d\lambda
\sum_{a}aP_{a}\left( \lambda ,A\right) \sum_{b}bP_{b}\left( \lambda
,B\right) .  \label{7}
\end{equation}
We might say that non-contextual theories assume that bodies possess
properties independent of any measurement (such properties, represented by $%
\lambda ,$ are not the observable quantities, but fully determine them).
However eq.$\left( \ref{7}\right) $ involves an assumption far stronger than
that, namely that the measurement of property $A$ is not perturbed by the
simultaneous measurement of $B$.

John Bell\cite{Bell} introduced local $HV$ theories, which are partially
non-contextual. Indeed these theories are non-contextual only for
measurements performed at spacelike separation, in the sense of relativity
theory. This implies that the expectation values of the products should be
calculated via eq.$\left( \ref{7}\right) $ when the measurements of $A$ and $%
B$ are performed at spacelike separation and via eq.$\left( \ref{6}\right) $
if this condition does not hold true. Spacelike separation is a rather
stringent condition. If the measurement of $A$ takes a time between $t$ and $%
t+\Delta t_{a}$ and that of $B$ between $t$ and $t+\Delta t_{b},$ all times
defined in an appropriate inertial frame, then spacelike separation requires
that 
\[
\max \{\Delta t_{a},\Delta t_{b}\}<d, 
\]
$d$ being the maximal distance between the two measuring equipments (\emph{d}
defined in the same frame).

In summary there are three kinds of HV theories fulfilling the following
relations of inclusion 
\[
general\supset local\supset non-contextual. 
\]
General HV theories are compatible with quantum mechanics and with
experiments as said above. The compatibility of local and non-contextual
ones is analyzed in the following.

\subsection{Non-contextual and local hidden variables vs. quantum mechanics}

Non-contextual HV theories are not compatible with quantum mechanics, a
statement known as Kochen-Specker theorem, although Bell proved it
independently\cite{BellRMP}, \cite{Mermin}. A proof goes as follows. We
consider four dichotomic observables $A_{1},B_{1},A_{2},B_{2},$ that is
observables which may take only the values $\pm 1.$ Calculating the
expectations via eq.$\left( \ref{7}\right) $ and making use of the
properties of probabilities (i. e. $0\leq P\leq 1)$ it is possible to derive
the inequality 
\begin{equation}
\left| \left\langle A_{1}B_{1}\right\rangle +\left\langle
A_{2}B_{1}\right\rangle +\left\langle A_{2}B_{2}\right\rangle -\left\langle
A_{1}B_{2}\right\rangle \right| \leq 2,  \label{8}
\end{equation}
which is the most popular form of a Bell inequality\cite{CHSH}. Eq.$\left( 
\ref{8}\right) $ is violated by the quantum predictions in some cases, for
instance in a system of two spin 1/2 particles (say silver atoms) in the
singlet spin state, flying in opposite directions, say along the axis $OZ$.
Measuring the spin projections on two different directions in the $XY$ plane
(say forming angles $\phi _{A}$ and $\phi _{B}$ with the $OX$ axis) the
quantum prediction for the correlation (as defined in eq.$\left( \ref{7}%
\right) )$ is 
\begin{equation}
\left\langle AB\right\rangle =\frac{1}{2}\left[ 1+\cos \left( \phi _{A}-\phi
_{B}\right) \right] .  \label{9}
\end{equation}
Hence choosing measurements (in different runs of the experiment) with four
different positions of the spin analyzers with angles 
\begin{equation}
\phi _{A1}=0,\phi _{B1}=\pi /4,\phi _{A2}=\pi /2,\phi _{B2}=3\pi /4,
\label{9a}
\end{equation}
the correlations predicted by quantum mechanics are 
\begin{equation}
\left\langle A_{1}B_{1}\right\rangle =\left\langle A_{2}B_{1}\right\rangle
=\left\langle A_{2}B_{2}\right\rangle =\frac{1}{2}+\frac{\sqrt{2}}{4}%
,\left\langle A_{1}B_{2}\right\rangle =\frac{1}{2}-\frac{\sqrt{2}}{4},
\label{10}
\end{equation}
which violates the Bell inequality $\left( \ref{8}\right) .$ This ends the
proof.

Local theories are incompatible with \emph{standard} quantum mechanics,
which is known as \emph{Bell's theorem}. (The word \emph{standard} in this
context will be explained in more detail below). The proof is quite similar
to the one for non-contextual theories, with the additional condition that
each of the four measurements involved in eq.$\left( \ref{8}\right) $ is
performed at spacelike separation. This similarity has been the source of
some confussion because some people have believed that any violation of a
Bell inequality refutes local HV theories, whilst the truth is that the
violation may only refute non-contextual HV theories (that is if
measurements are not performed at space-like separation.)

The physical model presented (or rather sketched) in this paper does not
pretend to be a new theory different from quantum mechanics, but rather an
interpretation of the quantum formalism. On the other hand the model may be
qualified as a HV model, as explained in the following. Consequently it is
crucial for the model to know whether it is local because in this case it
seems not to be a valid model, in view of Bell's theorem.

The main point of the model, as explained above, is the belief that quantum
vacuum fluctuations are \emph{real} fields. As a consequence the fields at
every spacetime point, or the coefficients of their Fourier analysis (like $%
\{c_{j}\}$\textbf{\ }and\textbf{\ }$\mathbf{\{}c_{j}^{*}\}$ in eq.$\left( 
\ref{2.1}\right) )$ may be taken as the hidden variables of the model. They
may be named hidden because they do not appear explicitly in the quantum
formalism. On the other hand the model may be said contextual as may be seen
with reference to eq.$\left( \ref{6}\right) .$ Indeed the vacuum
fluctuations (ZPF) may be influenced by the measurement context, and
therefore some correlation between the apparatus measuring $A$ and the
apparatus measuring $B$ may be established via the ZPF. Actually the
modification of the ZPF by material systems is well known. For instace the
Casimir effect is due to the modification of the vacuum fields by the
presence of metallic plates, as explained in section 2.2.

However the model, although contextual, is local. In fact the propagation of
the vacuum fiels is causal in the sense of relativity theory (as shown for
instance by eq.$\left( \ref{2.1}\right) $). Relativistic causality is what
Bell called locality. As a consequence it seems that the model is untenable
as an interpretation of quantum mechanics. My rebuttal to this statement is
the conjecture that \emph{experiments showing a violation of locality are
not feasible}, and quantum predictions do not (or should not) exist for
impossible experiments. The conjecture is reinforced by the history of the
attempts at refuting local HV theories via the empirical violation of a Bell
inequaltiy. In the next section I shall comment on these attempts, but here
I give an example showing the difficulties to perform a real test of local
hidden variables theories. So far there has been only one proposal for an
experiment able to measure the correlations involved in eq.$\left( \ref{10}%
\right) $ between the spin projections of two atoms\cite{Fry}. The
experiment is extremely involved as is shown by the detailed proposal
published in 1995, and the experiment has never been performed.

\subsection{Empirical tests of Bell's inequalities}

During the last 40 years many experiments have been performed with the
purpose of testing Bell\'{}s inequalities\cite{Genovese}, but only two kinds
of experiments will be commented here, involving either optical photon pairs
or entangled atoms. Less significance has been attributed to the remaining
experiments.

Until about 1983 optical tests involved entangled photon pairs produced in
atomic cascades, the Aspect\cite{Aspect} experiment being the most
celebrated of that period. However the authors were aware that the
experiments could not really violate the Bell inequalities because the
results could not fit the ideal quantum prediction 
\begin{equation}
\left\langle AB\right\rangle =\frac{1}{2}\left[ 1+\cos \left( 2\phi
_{A}-2\phi _{B}\right) \right] .  \label{10a}
\end{equation}
(In comparison with eq.$\left( \ref{10}\right) $ a factor 2 appears due to
the fact that photons have spin 1 rather than 1/2). In fact assuming for
simplicity a linear detection probabilty and labelling $\eta $ the detection
efficiency and $\varepsilon $ the noise to signal ratio, that is the ratio
between dark counts and true photon detections, the quantum prediction with
real detectors departs from eq.$\left( \ref{10a}\right) $ becoming 
\begin{equation}
\left\langle AB\right\rangle =1-\eta \left( 1+\varepsilon \right) +\frac{1}{2%
}\eta ^{2}\left[ 1+\cos \left( 2\phi _{A}-2\phi _{B}\right) \right] ,
\label{11}
\end{equation}
where I have neglected the noise in the coincidence detection (false
coincident counts). If we insert the result in the Bell inequality $\left( 
\ref{8}\right) $ we get 
\begin{equation}
\left| \left\langle A_{1}B_{1}\right\rangle +\left\langle
A_{2}B_{1}\right\rangle +\left\langle A_{2}B_{2}\right\rangle -\left\langle
A_{1}B_{2}\right\rangle \right| \leq \left| 2-2\eta \left( 1+\varepsilon
\right) +\eta ^{2}\left[ 1+\sqrt{2}\right] \right| \leq 2,  \label{12}
\end{equation}
where I have taken into account that the left hand side achieves the maximum
value if we choose the angles to be half those of eq.$\left( \ref{9a}\right) 
$. Therefore the Bell inequality is fulfilled whenever the following
inequality holds true 
\[
\eta <0.82\left( 1+\varepsilon \right) , 
\]
an inequality fulfilled in all performed optical experiments in the
commented period because available detectors had efficiencies below 10\%.
This difficulty for the tests has been called the \emph{detection loophole.}

In order to overcome the detection loophole\emph{\ }an assumption named 
\emph{no-enhancement} was introduced. It may be stated saying that \emph{in
any HV model for the experiments }(involving measuring the polarization
correlation of optical photons)\emph{\ the detection probability of a photon
cannot increase by crossing a polarizer. }Adding this condition to those
used by Bell in his derivation of the inequalities, it is possible to derive
new inequalities which were actually violated in the commented (atomic
cascade) photon experiments. Of course the HV models refuted were only those
fulfilling the no-enhancement assumption. The model presented in this paper
does not fulfil that condition, as may be seen by looking at eq.$\left( \ref
{2.2}\right) ,$ which represents the light intensity at the outgoing channel
of a \emph{balanced} beam splitter. It may be realized that the outgoing
intensity can be greater than the incoming one that is $\left|
E_{signal}+E_{ZPF}\right| ^{2}.$ The action of a \emph{polarizing} beam
splitter is more involved, but again the intensity of signal plus ZPF may
increase, whence the detection probability can be enhanced.

There is another difficulty with the empirical tests of Bell inequalities
involving photon pairs produced in atomic cascades. In fact the angular
correlation of the photon produced is too low to violate a Bell inequality
even if ideal, 100\% efficient, detectors were used\cite{Santosprl}. This is
due to the three-body character of the two-photon emission in atomic
cascades. In spite of these problems the experiments, in particular
Aspect\'{}s\cite{Aspect}, is currently quoted in books of quantum mechanics
as refuting local HV theories.

During the last decades, entangled photon pairs produced via parametric down
conversion are substitutes for those coming from atomic cascades. In this
case the problem of low angular correlation does not appear, but the
detection loophole remains. The additional assumption of no-enhancement has
been abandoned and a hypothesis of \emph{fair sampling }is used instead.
Thus it is assumed that \emph{the ensemble of photons actually detected is
representative of the full set of photons arriving at the detectors}. The
assumption allows extrapolating the actual experimental results, in
agreement with the quantum prediction eq.$\left( \ref{11}\right) ,$ to ideal
detection where $\eta =1,\varepsilon =0,$ which would violate the Bell
inequality. It is obvious that the violation only refutes the family of
local HV models fulfilling the fair sampling assumption. The question is
whether there are interesting local HV models not fulfilling \emph{fair
sampling}, and the answer is affirmative. Furthermore HV models not
fulfilling fair sampling are most natural. In fact, the purpose of hidden
variables is to explain why systems, represented by the same quantum state,
behave differently. In particular why some photons are more likely detected
than other photons, in spite of being treated as identical in the quantum
formalism. Therefore it is quite consistent with the idea of hidden
variables to assume that the photons actually detected belong to a special
subset of photons, that is those which are detectable with a probability
greater than the average. In such HV models the sample of detected signals
is not representative of the full set. The point may be more clearly
understood with refrence to our model, as explained in the following.

The model presented in this paper fits in a statement by Willis Lamb\cite
{Lamb}: ``Photons are the quanta of light, but they are not particles''. For
us photons are fluctuations of the electromagnetic field, maybe in the form
of needles of radiation, superimposed to the ZPF. As a consequence even the
sentence ``one hundred detection efficiency'' is meaningless in our model.
Detection efficiency might be defined only as the ratio between the number
of photocounts, excluding (an unknown number of) dark counts, and the
radiation energy arriving at the detector after subraction of the ZPF, the
energy being measured in units $h\nu .$ If the photon counters used in an
experiment have a relatively low detection efficiency, it is most natural to
assume that signals having higher intensity will be more likely detected.
Thus the sample detected is not representative of the whole set of arriving
signals, just because the signals are not identical. And the lack of real
equality of the signals is \emph{the crucial assumption} of HV models. In
summary, the \emph{detection loophole} is not a purely technical problem in
the manufacture of detectors, but derives from fundamental reasons according
to our model.

In experimental tests of a Bell inequality involving atoms the detection may
be quite efficient and the property corresponding to the polarization of
photons (i. e. a linear combination of different atomic states) has been
also measured with good efficiency in the experiment by Rowe et al.\cite
{Rowe}. Thus the Bell inequality $\left( \ref{8}\right) $ has been violated,
although some uncertainty exists about the statistical significance of the
violation\cite{SPRA}. As a consequence \emph{the experiment has refuted
non-contextual hidden variables theories} (see section 2 for the proof that
the violation of a Bell inequality refutes non-contextual HV theories).
However the measurements have not been made insuring spacelike separation,
and therefore local HV theories have not been refuted. A popular, but
misleading, form of the latter statement is to say that ``local hidden
variables theories have been refuted by the experiment, modulo the locality
loophole''. The correct statement is to say that the question whether local
HV models are possible is still open. My conjecture, already stated seven
years ago\cite{Sphil}, is that \emph{local HV models are possible, }that is%
\emph{\ local realism is compatible with experiments. }Furthermore I think
that\emph{\ it is also compatible with quantum mechanics, }provided that we
call quantum predictions only those for real experiments, that is excluding
``predictions'' for ideal, not feasible, experiments.

\subsection{From ideal to real experiments}

Asides from the possible difficulties posed by Bell\'{}s theorem, discussed
in the previous subsections, there are other reasons making our model
apparently untenable, which any well informed quantum physicists could
discover. It is impossible to rebutte all possible objection in a single
article of limited length and I shall give here only a few general
arguments. They may be summarized in the following three points.

1. Quantum mechanics itself introduces many constraints for real experiments
which are not taken into account in ideal examples. To mention just one I
shall consider the effect of the Heisenberg uncertainty relations for the
measurement of the spin correlation of two massive particles like electrons,
neutrons or atoms. Let us assume that both particles start at a point and
travel a distance $l$, moving in opposite directions with velocity $v$. If
there is an initial uncertainty in the position, $\Delta x,$ and the
velocity, $\Delta v,$ of one ot the particles, the length traveled by it
will have an uncertainty 
\[
\Delta l\simeq \Delta x+\frac{l}{v}\Delta v\geq \Delta x+\frac{l 
\rlap{\protect\rule[1.1ex]{.325em}{.1ex}}h%
}{2mv\Delta x}\geq 2\sqrt{\frac{l
\rlap{\protect\rule[1.1ex]{.325em}{.1ex}}h%
}{2mv}}, 
\]
$m$ bing the mass of the particle. The former inequality follows from the
Heisenberg uncertainty relation and the second one from trivial algebra. If
we want to measure the spins of the particles at a spacelike separation we
should insure that the uncertainty in their mutual distance should fulfil 
\[
\frac{2\Delta l}{v}\leq \frac{2l}{c}\Rightarrow l\gtrsim \frac{%
\rlap{\protect\rule[1.1ex]{.325em}{.1ex}}h%
}{mc}\left( \frac{c}{v}\right) ^{3}, 
\]
$c$ being the speed of light. This means that ``locality cannot be
violated'' until the particles have traveled some distance, macroscopic for
plausible values of $m$ and $v$. This does not put unsurmontable
difficulties for a (loophole-free) test of Bell\'{}s inequality but it is an
indication that the experiment is less simple than it might appear.

2. There may be predictions of the quantum formalism which seem to
contradict our model. An example, mentioned in subsection 2.4, is the fact
that the formalism predicts no rotational zeropoint. That is a system with
total angular momentum $\mathbf{J}$\emph{=0} does not rotate. I believe that
this quantum result should not be interpreted as the absence of any
rotation, but rather as the impossibility to determine the actual
fluctuating rotation in real experiments.

3. One of the cherised statements of quantum mechanics is the principle of
superposition. It says that if two states $\psi $ and $\phi $ of a system
are possible, then any linear combination of $\psi $ and $\phi $ represents
other possible state. Of course, the principle is limited by the
superselection rules, so that for instance the linear combinations are not
realizable states if $\psi $ and $\phi $ corresponds to states with
different electric charge. But I think that stronger constraints should
exist, that is I do not think that neither all vectors of the Hilbert state
may correspond to physically realizable states nor all selfadjoint operators
correspond to quantities actually measurable. However I am not in a position
to make a more precise statement.

In summary I believe that constraints on the quantum formalism may exist,
able to eliminate all apparent difficulties for the physical model of the
microworld here presented.

\section{Conclusions}

An intuitive picture of the quantum world would be useful and possible. The
starting point for that picture is to assume that quantum mechanics is a
stochastic theory and that typically quantum phenomena are due to an
universal noise in the form of \emph{real} vacuum fluctuations of all
fundamental fields present in nature.

An attempt at explaining every quantum phenomena from the vacuum
fluctuations with a clear and consistent model seems formidable. A better
approach would be to try to get a picture derived from the quantum
formalism, in particular via an interpretation of the commutation and
anticommutation rules, which I believe represent a characterization of the
randomness associated with quantum mechanics. An attempt in this direction
has already been made\cite{Comm}, but a deep understanding is still lacking.

A problem for viewing quantum mechanics as a stochastic theory, along the
lines here presented, is the alleged violation of the Bell inequalities. It
is the case that a loophole-free violation has not yet been produced in
spite of the big effort of many people during more than 40 years. For me
this failure is an indication that quantum mechanics is compatible with
local hidden variables \emph{for real experiments}, even if some $\emph{ideal%
}$ $\emph{(}$\emph{gedanken) experiments} violate the Bell inequality\cite
{Sphil}.

\end{document}